\documentclass{astronomyletters}

\usepackage{graphicx}
\usepackage{latexsym,amssymb,amsmath,color}
\usepackage{hyperref}
\usepackage{natbib}
\setcitestyle{aysep={,}}
\usepackage[utf8]{inputenc}

\begin{document}
	
\newcommand{\gcc}{\mbox{g/cm$^{3}$}}
\newcommand{\XC}{X_\text{C}}
\newcommand{\rhoi}{\rho_\mathrm{ign}}
\newcommand{\rhom}{\rho_{9}}
\newcommand{\rhoim}{\rho_\mathrm{ign,9}}
\newcommand{\rhomin}{\rho_\mathrm{min}}
\newcommand{\tdiff}{t_\text{diff}}
\newcommand{\tdiffa}{t_\text{diff1}}
\newcommand{\tdiffb}{t_\text{diff2}}
\newcommand{\tnu}{t_{\nu}}
\newcommand{\ttrans}{t_\text{trans}}
\newcommand{\Ttrans}{T_\text{trans}}
\newcommand{\Tmax}{T_\mathrm{max}}
\newcommand{\Tign}{T_\mathrm{ign}}
\newcommand{\Etot}{E_\mathrm{tot}}
\newcommand{\Egam}{E_{\gamma}}
\newcommand{\Enu}{E_{\nu}}
\newcommand{\Qb}{Q_\mathrm{b}}
\newcommand{\msun}{\mbox{${M}_\odot$}}
\def\etal{{et~al.}}

\def\received{ }      
\def\journame{ASTRONOMY LETTERS ~ Vol.~50 ~ No.~12 ~ 2024}

\authors[Ryabinkov,  Kaminker]{
   \nextauth{A.I. Ryabinkov}{1}      
  \nextauth[kam@astro.ioffe.ru]{A. D. Kaminker}{1}, 
}

\titles[spatial distribution of clusters]{Features of the spatial distribution of  Galaxy Clusters }

\affiliations{
\nextaffil{Ioffe Institute, Politekhnicheskaya 26, St.~Petersburg, 194021 Russia}
}

\wideabstract{
A statistical analysis of anisotropic quasiperiodic  features of the spatial distribution of galaxy clusters 
obtained on the basis of spectroscopic  and photometric redshifts in the interval  $0.1 \leq z \leq 0.47$  
has been carried out.    Based on  data from  the SDSS~III catalog, we show  that the 
preferential direction  previously detected in the  northern  hemisphere  (a narrow cone of directions: 
$\alpha_0=170^\circ \pm 5^\circ, \  \delta_0= 29^\circ \pm 5^\circ$),   along which the one-dimensional 
distribution of projections of  the  Cartesian coordinates of clusters contains a
significant  ($\gtrsim (4 - 5) \sigma$)   quasi-periodic component, can also be found using
photometric redshifts,  achieving a certain accuracy   ($\Delta z  \lesssim 0.013$).  
Based on  data from the photometric   DES$\times$unWISE   catalog,     
we have analyzed  the spatial distribution of clusters 
in the  southern hemisphere, where a cone of close
directions was also detected    ($\alpha_0=346^\circ \pm 5^\circ,\  
\delta_0= - 29^\circ \pm 5^\circ $), which are approximately 
an extention  of the directions   revealed in the  northern hemisphere.
The power spectra of one-dimensional distributions along these directions contain significant
($\gtrsim (4 - 5)  \sigma$) features  in the same  interval of wave numbers
$0.04 \lesssim k   \lesssim 0.06~h$~Mpc$^{-1}$. 
\doi{10.1134/S1063773725700124}  
}


\section{Introduction}
\label{s:introduct}

By now, 
several pieces of evidences
that the spatial distribution of cosmologically distant objects 
(e.g., galaxies or galaxy clusters) 
can
exhibit elements of
large-scale regularity (see, e.g., 
Broadhurst et al. 1990;
Szalay et al. 1993; Landy et al. 1996; Einasto et al. 1997b, 2011, 2016;
Saar et al. 2002; Einasto, 2014),
characterized by a 
fairly
wide range of scales
$(100 - 140)~h^{-1}$~Mpc,
have been accumulated in the literature.    
It should be noted that all such 
pieces of 
evidence  are  probabilistic  
in nature and correspond  to a relatively low  significance
level ($ \sim 3 \sigma$).  

In addition, a number of cited papers (see, e.g., Broadhurst et al. 1990;
Szalay et al. 1993; Landy et al.  1996) show that large-scale
quasi-periodicity can have an anisotropic character, i.e.,
manifest itself only in certain directions in space.
Such anisotropy can be characteristic of the quasi-periodic
anomalies manifested in the spatial distribution of
cosmological objects, and requires a special technique for its reliable
detection (e.g.,  Saar et al.  2002).

As a variation of this technique, in the work of
Ryabinkov  and  Kaminker (2021) 
proposed 
a method of projections of Cartesian coordinates of 
a  set  of objects under consideration
in the  comoving coordinate system (CS)
onto the $X$ axis of a rotating system,
whose center
is aligned with the center of the fixed  CS. 
In this case, the $X$ axis scanned
certain areas of the  northern hemisphere.
Based on the one-dimensional (1D) distributions
of  the projections, 
1D power spectra (PSs) $P_X(k)$ were calculated, and the significance 
level of individual PS peaks was determined for different $k$. 
By this method,
using data from the SDSS~DR7 galaxy catalog
in the range of spectroscopic redshifts 
$0.16 \leq z \leq 0.47$, the direction $X_0$ was found,
for which the peaks in the one-dimensional PS have the largest amplitude
(significance\  $\gtrsim (4 - 5) \sigma $).

The equatorial coordinates (EC)  of this direction were
$\alpha_0 \simeq 175^\circ - 177^\circ$  
and $\delta_0 \simeq 22^\circ - 27^\circ$, 
where $\alpha\ - $ right ascension and
$\delta\ -$  declination.
The peaks in the 1D PS  indicated the existence
in the specified $z$ interval
of an anisotropic quasi-periodic
component with a characteristic scale (quasi-period)
$\Delta X = 116 \pm 10~h^{-1}$~Mpc.

In the subsequent work (Ryabinkov, Kaminker, 2024)
two versions of the proposed
method of statistical analysis of the distribution of
1D projections of the Cartesian coordinates of galaxies
on different $X$ axes were used. 
The results of 
Ryabinkov and  Kaminker (2021)
were confirmed based on the
extended SDSS DR12 LOWZ galaxy catalog.
It was shown, in particular, that
in the same interval of spectrometric $z$
and approximately in the same direction
$X_0$ the 1D distributions of the projections
of the galaxies' coordinates contain
a significant ($ \gtrsim (4 - 5) \sigma$)
quasi-periodic component with the same characteristic
scale.

In the same work, a preliminary
analysis of the data from the  {southern} sky catalog
(Wen, Han, 2022) with the photometric
redshift data of galaxy   {clusters} was carried out.
As a result,
in the one-dimensional distribution of projections onto the $X_0$
axis of the {southern} hemisphere,
approximately opposite to the $X_0$
axis of the  {northern} hemisphere,
a quasi-periodic component was also detected,
the characteristic scale of which  $130 \pm 10~h^{-1}$~Mpc, 
taking into account the error,  is consistent  with the
scale of the  northern hemisphere discussed above.

In this paper, we consider galaxy clusters 
in both the  northern  and  southern hemispheres
in the redshift range of $0.1 \leq z \leq 0.47$.
Using
the technique developed in previous works,
we compared the spectroscopic $z$ determined for galaxy 
clusters with the photometric ones 
from the SDSS III catalog for the
northern hemisphere (Wen et al.  2012).
In addition,
based on the data from 
the catalog of
the Wen and Han (2022) catalog,
the photometric $z$ of clusters in the  southern 
hemisphere were considered. 
The existence of a single anisotropic
quasiperiodic anomaly with characteristic scales lying in the range of 
$100 - 140~h^{-1}$~Mpc is confirmed.

In Section~2 we describe the observational data studied in the work.
In Section~3 we introduce the main definitions
and basic quantities used in the further analysis.
In Section~4 we consider the analysis methodology and its results
as applied to the  northern  hemisphere, 
we compare the power spectra of the distributions of galaxy clusters
with the measured
spectroscopic and photometric
redshifts $z$. In Section~5 we analyze
clusters with photometric $z$
in the  southern  hemisphere.
Conclusions and discussions of the results are given in Section~6.

\section{OBSERVATIONAL DATA}
\label{sec:data}

In the  northern hemisphere, we use spectroscopic and photometric data
of galaxy clusters identified based on
the SDSS III galaxy catalog (Wen et al.  2012).%
\footnote{zmtt.bao.ac.cn/galaxy\_clusters/catalogs.html,\ ``WHL12''}
The ``WHL12'' catalog contains 42,929 clusters in the cosmological
redshift range of $0.1 \leq z \leq 0.47$.
The studied region of clusters
in the equatorial coordinate system ($\alpha$ and $\delta$)
is shown in the upper left part of Fig.~1a.
The uncertainty of spectroscopic redshifts is
$\delta z = \Delta z/(1+z) \sim (3 - 6) \times 10^{-4}$ 
(see, e.g., Bolton et al., 2012), which
is quite sufficient for reliable determination of inhomogeneities
(both irregular and quasi-regular) corresponding to the scales 
considered in this paper   $\Delta z \sim (3 - 5) \times 10^{-2}$.

In the  southern hemisphere, photometric data were used from
galaxy clusters identified using the extended ``DES$\times$unWISE'' 
galaxy catalog (Wen, Han, 2022).%
\footnote{zmtt.bao.ac.cn/galaxy\_clusters/catalogs.html,\ ``WH22''}
In the ``WH22'' catalog
in the redshift range of interest to us
$(0.1 \leq z \leq 0.47)$, 27,938 clusters were identified;
the average error in photometric $z$
is $\delta z \lesssim 0.013$. The region of clusters in the southern 
hemisphere is shown in the lower right part of Fig.~1a.

\begin{figure}[t]
\centering
\vspace{-3.5cm}
	\includegraphics[width=\columnwidth]{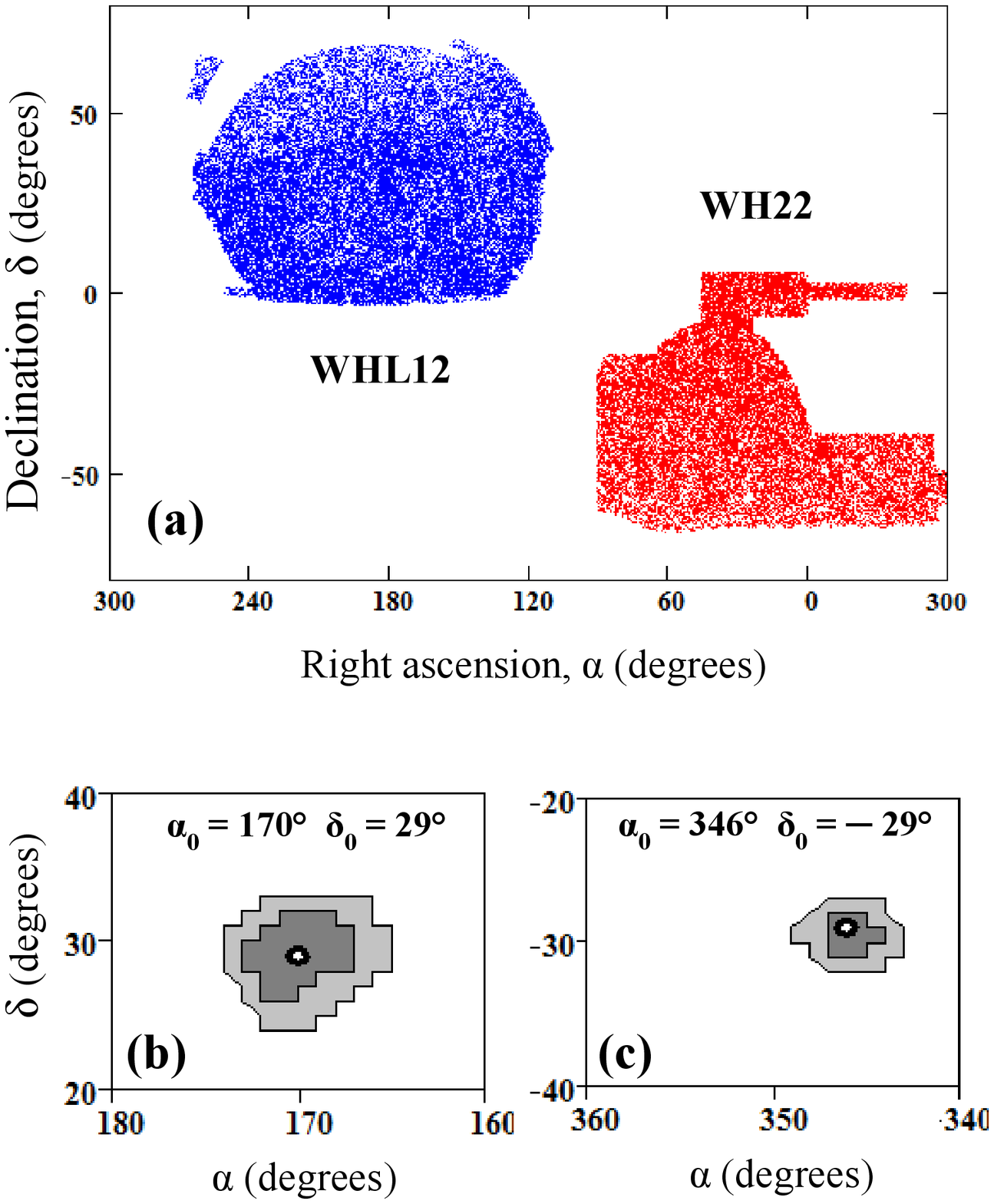} %
\caption{
Angular distributions of galaxy clusters considered in this paper 
in the equatorial coordinate system.
The  upper left  part of Fig.~1a
shows the region of clusters, mainly in the  northern hemisphere, 
identified  by  Wen et al.  (2012)
based on SDSS data (catalog ``WHL12'').
The  lower right  part  shows the region of clusters,
mainly in the  southern  hemisphere (catalog ``WHL22''), 
identified    by     Wen and  Han (2022)   based on photometric
data from the ``DES$\times$unWISE'' galaxy catalog. 
In Figs. 1b (in Section~\ref{sec:north} $-$ region (r1))
and 1c (in Section~\ref{sec:south} $-$ region (r3))
the confidence regions calculated  for
the highest peaks  in the 1D power spectra
$P_X (k)$   obtained
for  distributions of projections
of the cluster Cartesian coordinates onto the   $X$ axes 
of the cuboids at \
$k = k_\mathrm{max}=0.057~h$~Mpc$^{-1}$\  in  Fig.~1b\
and ~$k = k_\mathrm{max}=0.047~h$~Mpc$^{-1}$\   in  Fig.~1c,
respectively
(for more details, see Sections~\ref{sec:north} and \ref{sec:south});
gray colors correspond  to
confidence levels:  gray $-$ $3 \sigma$,  dark gray  $-$
$4 \sigma$,  black $-$ $5 \sigma$;
white stars in the centers  of confidence regions
indicate the directions   $X_0 ~(\alpha_0, ~\delta_0)$
of  the maximum  peak heights.
Figs.~1b and ~1c are bounded by rectangular
regions on the sky,  which   is scanned by the $X$ axis  when
calculating the significance levels presented in Figs.~3 and ~4.
} 
\label{fig1}
\end{figure}
  
Using the data of the specified catalogs,
we introduce a Cartesian coordinate system (CCS) 
for all galaxy clusters collected in them
in the comoving space.
In addition, a rotating CCS is introduced,
the center of which is coincides with the center of  the fixed CCS.
This allows us to study the features of the distribution 
of the projections of the
Cartesian coordinates of clusters on the $X$ axis
of the rotating CCS
and, thereby, to trace the dependence of these distributions
on the direction of the $X$ axis.

\section{BASIC  DEFINITIONS}
\label{sec:bd}

Let us consider the spatial distribution of galaxy clusters 
in a certain region of the sky using CCS:
\begin{eqnarray}
\label{XYZ}
&  & X_i = D (z_i) \sin(90^\circ - \delta_i) \cos \alpha_i  ,       \\
\nonumber  
&  & Y_i = D (z_i)  \sin(90^\circ - \delta_i) \sin \alpha_i ,         \\
\nonumber  
&  &  Z_i = D (z_i) \cos(90^\circ - \delta_i),
\end{eqnarray}
where $D(z_i)$ is the radial (along the line of sight) comoving
distance of the $i$-th cluster with redshift
$z_i$, $i=1,2,\ldots$ numbers the clusters (Kayser et al., 1997; Hogg, 1999)
\begin{equation}
D (z_i) = {c \over H_0}\, \int_0^{z_i} 
{1 \over \sqrt{\Omega_\mathrm{m} (1+z)^3 + 
\Omega_\Lambda}}\  \mathrm{d}z, 
\label{D}
\end{equation}
$H_0=100~h$~{km}~{c}$^{-1}$~{Mpc}$^{-1}$  $-$
is the Hubble constant in the modern era,
$c\ -$ is the speed of light;
$\alpha_i$ is the right ascension and $\delta_i$ is the declination
of the $i$-th cluster in the equatorial coordinate system (ECS).

In what follows, for comparison with our previous results (see above),
we use the same  $\Lambda$CDM model
$\Omega_\mathrm{m}=0.25$ and  
$\Omega_{\Lambda}=1-\Omega_\mathrm{m}=0.75$.

Following the works of
Ryabinkov and  Kaminker (2021, 2024)
we use two methods for constructing
1D distributions
of projections of the Cartesian coordinates of
clusters.

The first method  can be called the
\textit{Radon transform} method.
It is used in constructing Fig. 4b in Section~\ref{sec:south}. 
In essence, we are dealing
with a discrete analogue of the so-called $3D$
Radon transform (e.g., Deans, 2007),
applied to a certain sample
of galaxy clusters,
and we consider the projections onto the moving $X$ axis
of all Cartesian coordinates (in fixed CCS)
that fall into each  bin
(equal partition intervals)
within a given interval
$X_1 \leq X \leq X_2$.
Rotating the $X$-axis
and preserving the interval  ($ X_1,\ X_2$),
we successively project onto this axis the coordinates of all sample clusters
that fall within this interval.

Each fixed
direction of the $X$ axis is determined by the angles
$\alpha$ and $\delta$ of the ECS associated with the observer.
For each such direction,
an  1D  distribution of projections
on the $X$-axis of the coordinates of clusters 
belonging to the studied
spatial region is constructed,
and the  PS
of the obtained distribution is calculated. Thus,
the  height of the dominant peaks in each PS
at $k \sim (0.04 - 0.06) ~h$~Mpc$^{-1}$ and
the direction $X_0\ (\alpha_0,\ \delta_0)$,
along which the peak height is maximum
(see Fig.~1b,c) are found.

In these calculations, two main properties of the Radon transform take place:
1) \textit{translational invariance}, which
allows one to transfer the distribution of projections of the Cartesian
coordinates of objects obtained for a given $X$-axis
to any other $\tilde{X}$-axis parallel to the given one;
2) \textit{linearity}, which allows one to combine
projections of the coordinates of objects located in separate non-intersecting areas
into a single Radon transform of the entire sample as a whole.
Note that the  Radon transform method can only obtain the preferential 
direction (but not thenposition)
of the $X_0$-axis in space,
which can also correspond to any other $\tilde{X}_0$-axis.

The second method, which is a variation of the first one,
can be called the   \textit{scanning  (rotating) cuboid} method.
Below it is used to construct Fig.~3a, b in the Section~\ref{sec:north} 
and Fig. 4a in the Section~\ref{sec:south}. In these cases, we
construct a rectangular parallelepiped ({cuboid}) in space
with fixed coordinates of its vertices $X_1 \leq X \leq X_2,\
Y_1 \leq Y \leq Y_2,\ Z_1 \leq Z \leq Z_2$
in a moving CCS rigidly connected to the cuboid, 
while the direction of the $X$ axis
sets the orientation of the cuboid
in the  fixed  ECS.
Rotating the cuboid together with the $X$ axis
relative to the origin $X=0$ and defining the faces of the cuboid for each fixed
direction of the $X$ axis, we project onto this axis
the coordinates of all clusters from the given catalog,
located inside the cuboid.

As in the first method,
each fixed direction is determined by the angles
$\alpha$ and $\delta$ of the ECS,
and for each direction the PS is calculated.
With the obvious similarity of the proposed methods,
the difference of the {rotating cuboid} method is that
each direction of the $X$ axis is uniquely associated
with a certain sample of objects (clusters of galaxies) 
that fall inside the boundaries of the cuboid.
This allows us to approximately localize
the preferential  directions of the spatial distribution of objects,
detected at a certain level of significance.

Both methods (being essentially integral)
turn out to be quite sensitive
to the presence of rarefied, difficult to detect,
quasi-periodic components and the associated possible
anisotropy of the spatial distribution of objects.
In both approaches considered above, the basic
quantity for spectral analysis is the 1D-distribution function
$N_X (X)$, accumulating the projections of coordinates onto each
fixed $X$-axis,
$N_X (X) \mathrm{d}X$ $-$ the number of galaxy clusters
whose projections onto the $X$-axis
fall within the interval $\mathrm{d} X$.
Using the $N_X (X)$ function,  we calculate
the so-called normalized 1D-distribution function
within the framework of the bin approach
\begin{equation}
\mathrm{NN} (X_c^l) = {N_X (X_c^l) - S_X
\over \sqrt{S_X}},
\label{NNX}
\end{equation} 
where $X_c^l$ is the central point of the $l$-th bin,
$l=1,2,\ldots,{\cal N}_b$ numbers the bins,
$S_X$ is the average value of the function
$N_X (X_c^l)$ over all the bins under consideration. 
Note that the value NN$(X_c)$
can be considered as the signal-to-noise ratio
in the function of the values of $X_c^l$
along the selected axis.

The value NN$(X_c^l)$ allows
to calculate the 1D power spectrum (1D-PS).
\begin{eqnarray}
\label{PXk}
P_X(k_m) & = & |F_X (k_m) |^2 =   \\
\nonumber 
&  & {1 \over {\cal N}_b} \left\{ \left[ \sum_{\ l=1}^{{\cal N}_b} 
NN(X_c^l) \cos (k_m X_c^l) \right]^2 + \right. \\
\nonumber
& + & 
\left.  \left[ \sum_{\ l=1}^{{\cal N}_b} 
NN(X_c^l) \sin (k_m X_c^l) \right]^2  \right\},
\end{eqnarray}
where $F_X (k_m) =
({\cal N}_b)^{-1/2} \sum_{\ l=1}^{{\cal N}_b} NN(X_c^l)\ e^{ - i (k_m X_c^l)}$
$-$ one-dimensional discrete Fourier transform,
$k_m =2 \pi m / L_X$ -- wave number corresponding to
integer harmonic numbers $m=1,2,\ldots,{\cal M}$,\
the maximum number of independent harmonics
${\cal M}=\lfloor {\cal N}_b/2 \rfloor$
is determined by the so-called Nyquist number,
$\lfloor x \rfloor$ denotes
the largest integer $\leq x$, where $x$ is
an arbitrary real (positive)
number; $L_X$ is the full interval along the $X$-axis, 
the so-called  {sample length} in a given direction 
of the configuration space.
 
We rotate the $XYZ$ coordinates of the
moving CS
(in the second case, rigidly connected
to the cuboid) by certain Euler angles,
so that the new $X'$-axis%
\footnote{Here and below, the designation $X'$ instead of $X$
indicates the $X$ axis of the rotating CCS.}
is oriented in a certain direction
($\alpha'$ and $\delta'$) relative 
to the original (fixed) ECS.
Performing a sequence of such rotations,
we find the $X_0$ direction along which
the 1D-PS contains the highest peak,
at $k=k_\mathrm{max}$, lying in the interval
$0.04 \leq k \leq 0.06~h$~Mpc$^{-1}$.

For the homogeneity of statistical conditions
in different directions $X'$
we fix the same boundaries
of the scanning axes $X_1 \leq X' \leq X_2$, but different ones for
the northern and southern hemispheres,
as shown in Figs.~3a and 4a.
For example, for a {rotating cuboid} in the northern hemisphere,
we have $132 \leq X' \leq 1242~h^{-1}$~Mpc,
between which are located
${\cal N}_b = 111$ independent bins
with a width of $\Delta_{X} = 10~h^{-1}$~Mpc.%
\footnote{In this case, we select
such scanning areas by the $X'$-axis,
so that at the boundaries of these areas
all bins $\Delta_{X'}$
within the interval $X_1 \leq X' \leq X_2$
are filled with projections of the object coordinates 
approximately uniformly.}

To  estimate  the significance of the revealed  peaks in the PS, 
we use the exponential distribution of the heights (amplitudes) 
of the peaks $P_k$ in the power spectra
(e.g., Jenkins and  Watts 1968),
corresponding to the normally distributed random variables
Re$[F_X (k)]$ and Im$[ F_X (k)]$ with zero means
$\langle $Re$[F_X (k)] \rangle = \langle $Im$[F_X (k)] \rangle = 0$
and variances
$\sigma^2 (k) = \langle \{$Re$ [F_X (k)] \}^2 \rangle + \langle \{ $Im$ [F_X (k)] \}^2 \rangle =
\langle P_X (k) \rangle $,
where Re and Im $-$ are the real and imaginary parts
of the Fourier transform for a given harmonic $k$,
$P_X (k)$ is defined in (\ref{PXk}),
$\langle ... \rangle$ $-$ is the averaging over the statistical ensemble.
In this case,
the probability density function 
of the quantities $z = P_k$ is represented as
\begin{equation}
{\cal P}_X (z) \mathrm{d} z = 
{1 \over \sigma^{2} (k)}  \exp{(- z / \sigma^2 (k))} \mathrm{d} z.
\label{calP}
\end{equation}

Then the cumulative probability function
that the random peak height $P_k$
is less than some
fixed value $P_k < P^*_k$,
can be written
(see, e.g., Scargle, 1982; Feldman, 1994; Frescura, 2008):
\begin{equation}
{\cal F}(P_k < P^*_k,\  \lambda) = 1 - \exp(-\lambda\ \cdot P^*_k )
\, \, , \, \, \, \, \,  P^*_k  \geq  0,
\label{calF}
\end{equation}
where the exponential distribution parameter
$\lambda= \lambda (k) = \sigma^{-2}(k) = \langle P_X (k) \rangle^{-1}$
in contrast to the works cited above,
in which it was assumed that $\lambda =$ const,
is a relatively smooth function of $k$.
An essential condition for the validity of the given formulas
is the statistical independence of different $k$
(e.g., Lomb, 1976).
As a result, the expression ~(\ref{calF})
allows us to obtain the significance levels of the peaks $P_X (k)$,
if we calculate the averaged $\langle ... \rangle$
amplitudes of the PS for different $k$.

As such an averaging (over the ensemble), we average
PS  $\langle P_X (k) \rangle$ obtained
by the $X'$ axis scanning 
of certain regions in the sky
(see Sections~\ref{sec:north} and \ref{sec:south}).
In practice, for estimating the significance levels
using the expression~(\ref{calF})
it is convenient to use a fitting approximation,
the details of which are described, for example, in Ryabinkov, Kaminker (2021).
In this case, by virtue of the projection-slice theorem (see, e.g., Deans, 2007) 
we assume that the power spectrum of the 1D Radon transform averaged over a set of directions
$\lambda^{-1}(k) = \langle P_X (k) \rangle$
is approximated by the model function
\begin{equation}
f (k) = f_\mathrm{CDM} (k) + 1,
\label{f}
\end{equation}
where $f_\mathrm{CDM}(k)$ describes 3D fluctuations of the cold dark matter 
density (``CDM''), averaged over directions $k = |\vec{k}|$
(Bardeen et al., 1986), and
``1'' on the right-hand side of the equality ~(\ref{f})
describes the so-called {short noise}
(e.g., Feldman et al., 1994), 
uniformly distributed over all $k$ under consideration.

Using ~ (\ref{calF}) and
taking into account the approximate equality
$\lambda(k) \simeq f (k)^{-1}$,
we can construct smoothed
confidence probability levels
$\beta (k) = {\cal F}(P_k < P^*_k)$
for any $k $
in the entire interval ($0 \leq k \leq 0.3$).
In the following sections, we use
such smoothed dependences on $k$
to  estimate  the significance of the  peaks in 1D-PS
(see Figs. ~3 and 4).
At the same time,
due to the similarity of the two methods considered
above, we apply
the same approach  to  estimating  the significance of peaks 
in the PS obtained by the
{rotating cuboid} method.

In contrast to the works of Ryabinkov and Kaminker (2021, 2024), 
in this paper
we consider the same scanning areas
(see Sections~\ref{sec:north} and \ref{sec:south}) 
both for searching for
preferential
directions (near $X_0$) and for
assessing the significance of the obtained peaks.
This can lead to an underestimation
of the value of $\lambda (k)$ and, thus, to an overestimation
of the confidence level $\beta (k)$, i.e. an underestimation
of the significance of the revealed  peaks.

\section{NORTHERN HEMISPHERE. ROTATING CUBOID}
\label{sec:north}

In the  northern hemisphere,
the search for traces of quasi-periodicity in certain directions of
the spatial distribution of galaxy clusters was carried out
on the basis of the ``WHL12'' catalog (see Section~\ref{sec:data})
by the rotating cuboid method (see Section~\ref{sec:bd}).
To simplify the search procedure, we
take into account the results of the article by
Ryabinkov, Kaminker (2024)
obtained on the basis of statistical analysis 
of the SDSS DR12 LOWZ galaxy catalog.
We start the search from the same direction $X_0$
($\alpha_0 =170^\circ,\ \delta_0 = 28^\circ$) and for the same
$k=k_\mathrm{max}=0.054~h$~Mpc$^{-1}$,
as in the cited paper,
and obtain a peak in the power spectrum, but of significantly lower
significance $\sim (3 - 4) \sigma$\ (estimates of the
significance are discussed below).
Then the sizes of the cuboid faces are varied so that
the height of the peak in the power spectrum
for the same direction $X_0$
reaches the greatest value.
The coordinates of the vertices established in this way
are given in Fig.~3a.

Then, with a step of $1^\circ$, the direction of the
maximum peak $X_0$ is corrected and the value of 
$k_\mathrm{max}$ is refined,
which ultimately leads to the values of the
angles $\alpha_0 =170^\circ,\ \delta_0 = 29^\circ$
and the scale $2 \pi / k_\mathrm{max} = 111 \pm 10~h^{-1}$~Mpc.
Fixing the    derived  coordinates of the vertices
of the cuboid in its own coordinate system,
we scan with the $X$ axis
of the moving CS with a step of $1^\circ$
two rectangular regions in the $\alpha,\ \delta$ plane. 
These two scanning areas
are defined by the angle intervals:
$160^\circ \leq \alpha \leq 180^\circ$,
$20^\circ \leq \delta \leq 40^\circ$
$-$ area (r1)\ (see Fig.~1b)
and
$140^\circ \leq \alpha \leq 190^\circ$,
$10^\circ \leq \delta \leq 50^\circ$
$-$ area (r2).
Note that area (r2) entirely includes
the smaller area (r1).

\begin{figure}[t]
\centering
\vspace{-4.2cm}
\includegraphics[width=0.95\columnwidth]{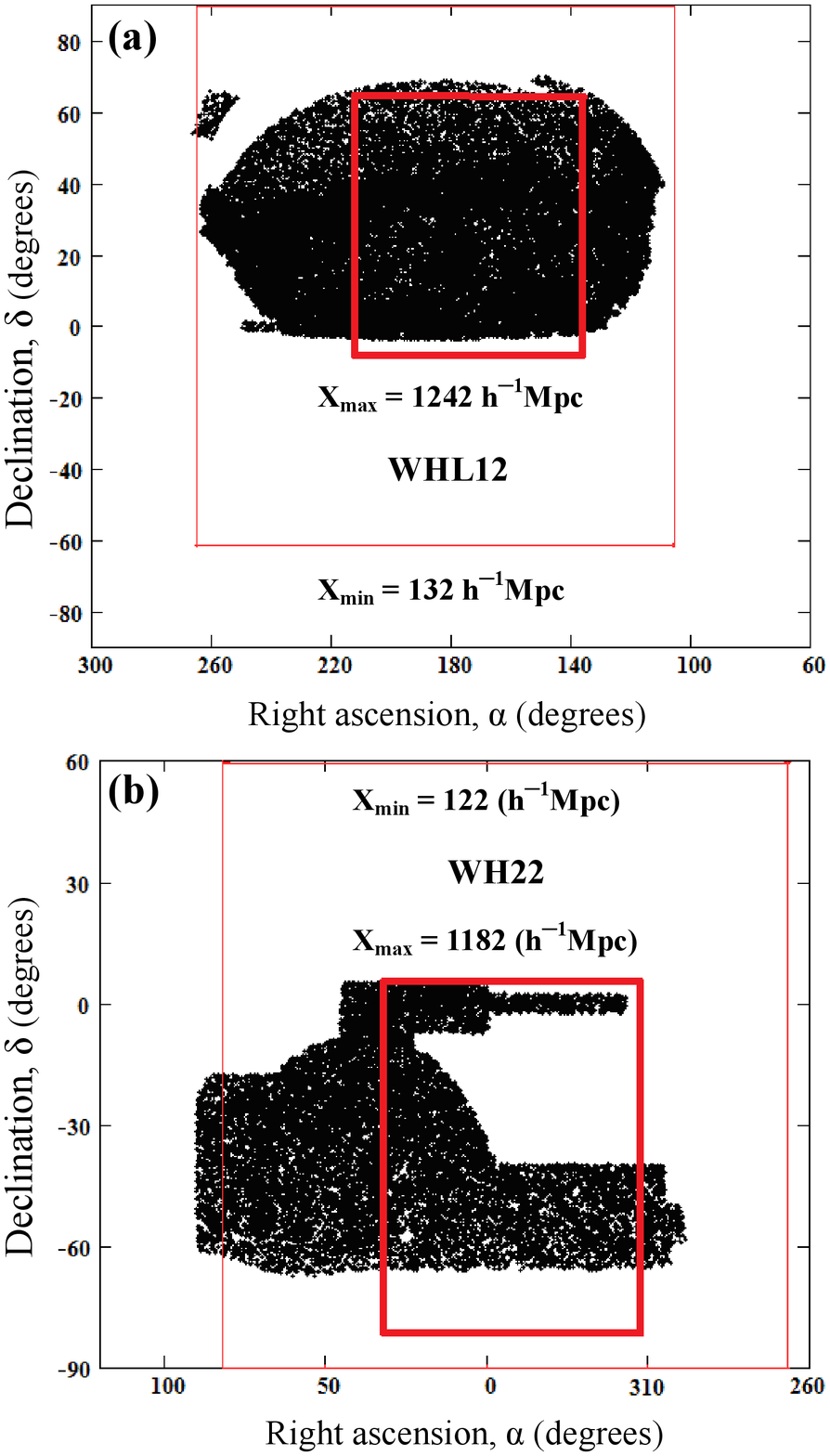} %
\caption{
The same angular distributions of galaxy clusters,
as in Fig.~1a, but with rectangular boundaries 
superimposed on them, which are
formed by the extreme positions of
two faces of cuboids corresponding to
the maximum $X_\mathrm{max}$ and minimum $X_\mathrm{min}$
when scanning by the $X$ axis
of the regions  (shown in Figs.~1b and ~1c)
in the  northern  $-$   Fig.~2a\ and
southern  $-$  Fig.~2b\  hemispheres, respectively;
 the faces that  being most distant from the origin of coordinates
(thick lines) correspond to the values
$X_\mathrm{max}=1242~h^{-1}$~Mpc (Fig.~2a) and
$X_\mathrm{max} = 1182~h^{-1}$~Mpc (Fig.~2b),
while   the faces that being  least distant
(thin lines)  correspond to the values 
$X_\mathrm{min} = 132~h^{-1}$~Mpc\ (Fig.~2a) and
$X_\mathrm{min} = 122~h^{-1}$~Mpc\ (Fig.~2b).
} 
\label{fig2}
\end{figure}

When scanning the region (r1) on the sky, shown in Fig.~1b, 
with the cuboid axis $X$, different sections of the cuboid, perpendicular
to the $X$ axis, outline different rectangular regions
on the plane ($\alpha,\ \delta$). For clarity, Fig.~2a
shows two such regions, corresponding to the extreme positions
of the cuboid faces when the $X$ axis rotates within the scanning region,
shown in Fig.~1b. The smaller rectangle, bounded by thick lines in Fig.~2a,
corresponds to the most distant
face of the cuboid $X_\mathrm{max}=1242~h^{-1}$~Mpc, and the larger one, 
bounded by thin lines, $-$ to the face closest to the origin
of the coordinates $X_\mathrm{min} = 132~h^{-1}$~Mpc. 
It is evident that even when scanning a limited area, shown in Fig. 1b, 
the cuboid in various positions captures majority of clusters  
of the considered part of the ``WHL12'' catalog.

As a result of scanning
the region (r1), one can calculate
1D power spectra $P_X(k)$
for each direction of the cuboid axis $X$ 
and after averaging $\langle ... \rangle$
obtain averaged
power spectra $\langle P_X (k) \rangle$
for any fixed $0 \leq k \leq 0.3$.
Using Eq.~(\ref{calF}), assuming that
$\lambda (k) = \langle P_X (k) \rangle^{-1}$,
one can construct relatively smooth curves
of significance levels,
shown in Fig.~3a and 3b.
On the other hand, as a result of the same scanning
of the region (r1)
we find directions of increased
significance (confidence regions) 
of the peaks $P_X (k)$
at a fixed $k=k_\mathrm{max} = 0.057~h$~Mpc$^{-1}$.
The confidence regions (including the direction $X_0$) 
obtained in this way
are colored in shades of gray in Fig.~1b.

\begin{figure}[t]
	\centering
\vspace{-0.3cm}
	\includegraphics[width=0.92\columnwidth]{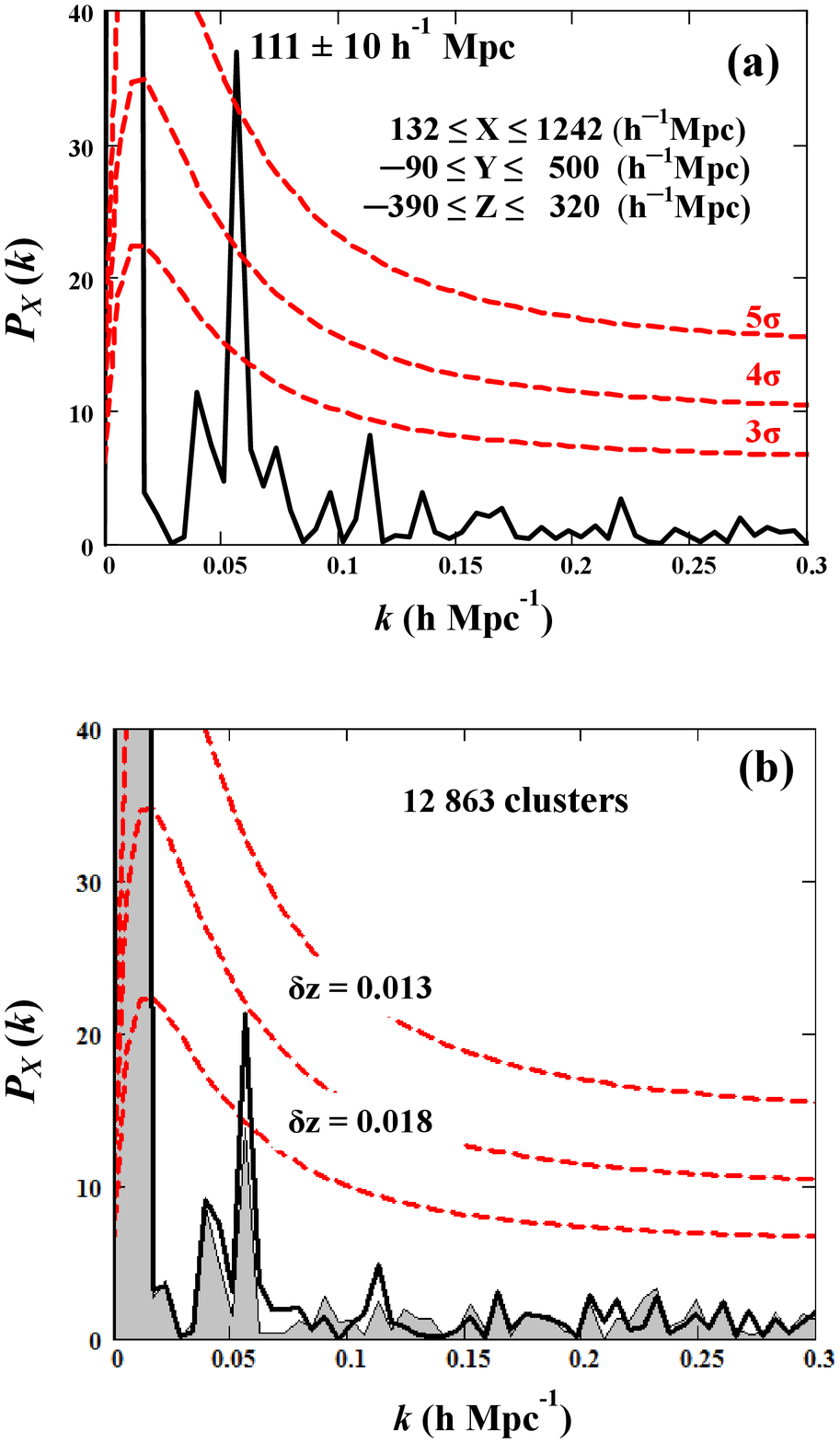} %
	\caption{
Power spectra $P_X(k)$ (solid lines) of the 1D distributions of the projections
onto  the $X_0$ axis (see below) of the Cartesian coordinates of galaxy clusters
of the  northern hemisphere 
falling into the cuboid bounded by the $X,\ Y,\ Z$ coordinates
(see the inset in Fig.~3a)
in its own reference frame;
the fixed direction
of the $X = X_0$ axis ($\alpha_0$ and $\delta_0$ see in Fig.~1b)
corresponds to the peak of maximum height
at $k=k_\mathrm{max}=0.057 \pm 0.005~h$~Mpc$^{-1}$\
or the scales
$111 \pm 10~h^{-1}$~Mpc. \\
Fig.\ 3a:\
Cartesian coordinates of clusters  calculated
only with \textit{spectroscopic} $z=z_\mathrm{sp}$;
dashed lines correspond to
significance levels
($ 3\sigma,\ 4\sigma\ and\ 5\sigma$)
calculated
(for all $k \leq 0.3~h$~Mpc$^{-1}$)
through  scanning by the rotating  $X$  axis  of
the rectangular region
shown in Fig.~1b  (see text). \\
Fig.~3b:\  Same as in Fig.~3a,
but with the  spectroscopic 
($z_\mathrm{sp}$) partly replaced by  photometric ($z_\mathrm{ph}$) for those
of the considered clusters for which one of the two conditions is satisfied:
$\delta z = | z_\mathrm{sp} - z_\mathrm{ph}|/(z_\mathrm{sp}+1) \leq 0.013$\ --\  thick  curve
or $ \delta z \leq 0.018$\ --\  thin curve bounding the region shaded in gray. 
The statistics of the clusters common to all three distributions
is  indicated.
}
\label{fig3}
\end{figure}

Scanning over a larger region (r2)
is performed to check for the absence of 
additional regions with increased
significance of peaks $\gtrsim 3 \sigma$
in the 1D  power spectra $P_X (k)$
at $0.04 \leq k \leq 0.06~h$~Mpc$^{-1}$.

The power spectrum $P_X (k)$ obtained 
for the direction $X_0$ is shown in Fig.~3a.
The spectrum is constructed taking into account the data
of 12,863 galaxy clusters
in the redshift range $0.1 \leq z \leq 0.47$, 
presented in the ``WHL12'' catalog
(see Section~\ref{sec:data}) and falling
into a cuboid with the $X = X_0$ axis and 
the dimensions indicated in the figure.
The Cartesian coordinates of the clusters
and their projections onto the $X_0$ axis
were calculated using formulas (\ref{XYZ})
taking into account only the {spectroscopic} $z=z_\mathrm{sp}$.
It is evident that the significance of the dominant peak 
in the power spectrum
exceeds $5 \sigma$. However, given the uncertainty 
of significance estimates
(see Ryabinkov, Kaminker, 2024), more cautious estimates 
$\gtrsim (4 - 5) \sigma$ are adopted here and below.
\footnote{In sections 4 and 5 of the RK24, 
the variances of significance levels where calculated  for 
smoothed power spectra of different realizations
obtained by successive shifts of angular lattices
(in $\alpha$ and $\delta$ coordinates)
corresponding to a 
weak correlations between  
directions of their nodes.}

Fig.~3b shows the power spectra for the same cluster statistics, 
but now a certain part of the {spectroscopic}
$z_\mathrm{sp}$ is replaced by photometric $z_\mathrm{ph}$, 
provided the deviation 
$\delta z = |z_\mathrm{sp} - z_\mathrm{ph}|/(z_\mathrm{sp} +1)$ 
does not exceed a given  value. 
Spectra are shown for two values $\delta z \leq 0.013$
(thick lines) and $\delta z \leq 0.018$ (thin lines,
bounding the regions shaded in gray).
For the first constraint, $63\%$ of the spectroscopic $z_\mathrm{sp}$
are replaced by photometric $z_\mathrm{ph}$, and for the second one this 
fraction  is already $76\%$.
 
If we assume that the values of $\delta z$ characterize the accuracy of
photometric measurements, then it follows from Fig.~3b that
the magnitude of the peak in the power spectrum depends significantly on this accuracy.
At $\delta z \leq 0.013$ the significance of the peak approaches $4 \sigma$,
and at $\delta z \leq 0.018$\ --\ $3 \sigma$.
It follows from this that
comparison of the height of the peak in the power spectrum
based on photometric measurements of $z$,
with the height of the peak obtained under the same conditions, 
but based on
spectroscopic measurements, with the necessary
confirmation and revision
can be used to test
the accuracy of photometric
measurements.

\section{SOUTHERN HEMISPHERE. ROTATING CUBOID AND RADON TRANSFORM}
\label{sec:south}

In the  southern hemisphere, the search for quasi-periodicity
in certain preferential  directions
of the spatial distribution of clusters
was carried out on the basis of data from the
``WH22'' catalog (see Section \ref{sec:data})
by two methods:
{scanning cuboid} (Fig.~4a),
and {Radon transforms} (Fig.~4b).

Using the {scanning cuboid} method, we select 
as the initial direction of the $X_0$ axis 
the direction opposite to that found for the {northern}
hemisphere: $\alpha=350^\circ,\ \delta=-29^\circ$.
Fixing this direction, we vary the coordinates of the cuboid 
vertices in its own coordinate system,
to obtain the greatest height of the peak corresponding to the same
$k=k_\mathrm{max} \simeq 0.057~h$~Mpc$^{-1}$,
as in the {northern} hemisphere.

Next, considering the derived  coordinates of the
vertices {of the rotating cuboid}
as a zero approximation, we successively vary the direction of the $X_0$ axis
with a step of $1^\circ$, and then
refining the coordinates of the cuboid vertices.
As a result, the coordinates of the $X_0$ axis in the ECS are determined as
$\alpha_0=346^\circ$ and $\delta_0 = -29$;  the final coordinates of the
vertices of the cuboid are shown in Fig.~4a.
In this case, the position of the peak $k=k_\mathrm{max}$ is 
shifted toward smaller values of $k$,
i.e. $k=k_\mathrm{max} \simeq 0.047~h$~Mpc$^{-1}$.
The coordinates of the $X_0$ axis in the ECS are shown in Fig.~1c, 
where the confidence regions (colored in shades of gray) are also shown 
for the directions corresponding to the highest peak  in the PS.

The confidence regions and the direction of the $X_0$ axis obtained in this way
coincide with the direction and confidence regions obtained 
by a similar variation procedure using the
{Radon transform} method  (Fig.~4b).
The optimal interval of
$X$ values in both cases
is equal to
$132 \leq X \leq 1172~h^{-1}$~Mpc, while
the wave number corresponding to the maximum peak in Fig.~4b is
$k=k_\mathrm{max} \simeq 0.048~h$~Mpc$^{-1}$.

Fixing the obtained coordinates of the vertices of the
cuboid in its own CS, we perform a sequential scan by the $X$ axis
with a step of $1^\circ$
of two rectangular regions in the $\alpha,\ \delta$ plane.
As in the {northern} hemisphere, both scanning regions
are located inside the rectangle,
bounded by thick lines in Fig.~2b, and
are determined by the angle intervals:
$340^\circ \leq \alpha \leq 360^\circ$,
$-40^\circ \leq \delta \leq -20^\circ$
$-$ region (r3)\ (see Fig.~1c)
and
$320^\circ \leq \alpha \leq 360^\circ$,
$-50^\circ \leq \delta \leq -10^\circ$
$-$ region (r4).
As in the  northern  hemisphere (Section \ref{sec:north})
the region (r4) entirely includes
the smaller region (r3).

\begin{figure}[t]
	\centering
	\includegraphics[width=\columnwidth]{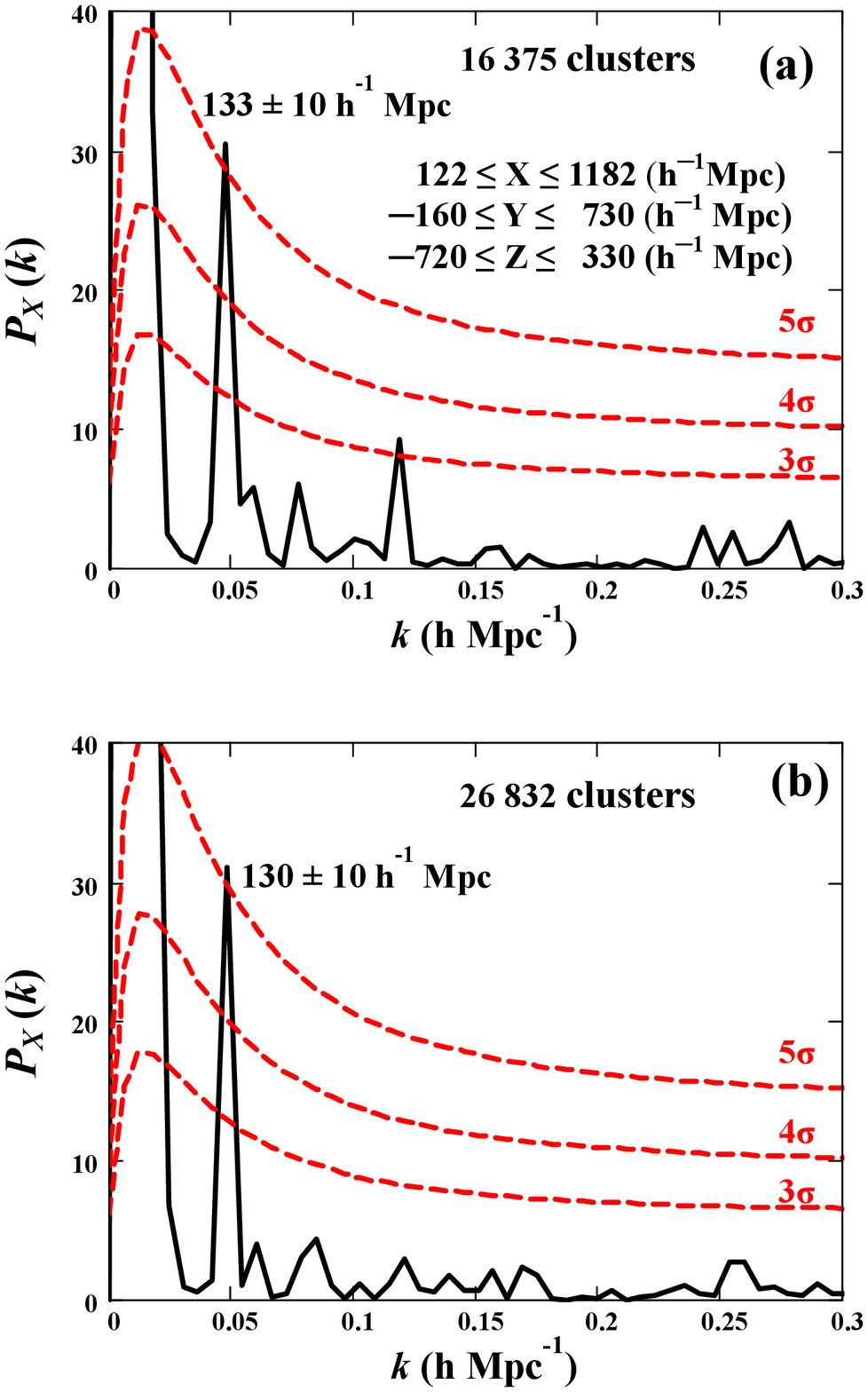}  %
	\caption{
Fig.~4a  is  organized  similarly to
Fig.~3a  but for the  set  of galaxy clusters
in  the  southern hemisphere
from the ``WH22'' catalog (see Section \ref{sec:data}) 
containing only  photometric $z_\mathrm{ph}$,  determined
with an average accuracy of $\delta z \sim 0.013$ 
over  the interval $0.1 \leq z \leq 0.47$;
the direction $X_0$ (shown in Fig.~1c)
corresponds to the peak of maximum height at
$k=k_\mathrm{max}=0.047 \pm 0.004~h$~Mpc$^{-1}$\
or the scales  $133 \pm 10~h^{-1}$~Mpc;
the coordinates  $X,\ Y,\ Z$   
bounding  the cuboid 
and statistics of clusters 
falling into the cuboid with the  $X=X_0$ axis 
are given in  insets;
dashed lines correspond to the significance levels
($ 3\sigma,\ 4\sigma\ and\ 5\sigma$), calculated
through  scanning  by the  rotating  $X$ axis  over   the rectangular region
shown in Fig.~1c.\\
Fig.~4b $-$ power spectrum $P_X(k)$ (solid lines)
of the 1D  distribution of projections 
onto the $X_0$ axis  of the Cartesian coordinates
of all clusters of the  southern  hemisphere,
falling into the interval $132 \leq X_0 \leq 1172~h^{-1}$~Mpc
(Radon transform, \ see  Section~\ref{sec:bd});
the peak of the maximum height   $(\gtrsim 5\sigma)$   corresponds to
$k=k_\mathrm{max}=0.048 \pm 0.004~h$~Mpc$^{-1}$\
or   the scales  $130 \pm 10~h^{-1}$~Mpc;
the accuracy of  photometric $z_\mathrm{ph}$ is the same
as in Fig.~4a.
The dashed lines show the significance levels,
calculated for the same rectangular region as in Fig.~4a,
but using the  Radon transform.
}
\label{fig4}
\end{figure}

By scanning the  region (r3)
and calculating the corresponding
1D\   PS $P_X(k)$
for each direction of the
$X'$ axis, we can obtain the averaged
PS $\langle P_X (k) \rangle$
for any fixed $0 \leq k \leq 0.3$.
As in Section~\ref{sec:north},
using the formulas (\ref{calF}) and
$\lambda (k) = \langle P_X (k) \rangle^{-1}$,
we can depict the
significance levels shown in Fig.~4a and 4b.

Scanning over a larger area (r4),
as in the {northern} hemisphere,
serves to check for the absence of
additional areas with increased
significance of the peaks ($\gtrsim 3 \sigma$)
in the  1D\   PS $P_X (k)$
at $0.04 \leq k \leq 0.06~h$~Mpc$^{-1}$.

As in Section~\ref{sec:north}, Fig.~2b shows for clarity 
two regions corresponding to the extreme positions of the cuboid faces 
when the $X$ axis rotates within the scanning area (Fig.~1c). 
The smaller rectangle, bounded by thick lines, corresponds to the most distant
face of the cuboid $X_\mathrm{max}=1182~h^{-1}$~Mpc, 
and the larger one, bounded by thin lines, $-$ the face closest to the origin
of the coordinates $X_\mathrm{min} = 122~h^{-1}$~Mpc.
We see that when scanning the cuboid in different positions
captures the vast  majority of the southern hemisphere clusters 
included in the catalog.

Fig.~4a shows the
power spectrum $P_X (k)$ obtained by the rotating cuboid method
with the highest peak height at
$k=k_\mathrm{max} \simeq 0.047 \pm 0.004~h$~Mpc$^{-1}$
(quasi-period $133 \pm 10~h^{-1}$~Mpc),
corresponding to the $X_0$ direction.
The spectrum is calculated taking into account the data
of 16\,375 clusters
in the redshift range $0.1 \leq z \leq 0.47$. 
Cartesian coordinates of clusters
and their projections onto the $X_0$ axis
were calculated using formulas (\ref{XYZ}) with
considering  only photometric $z=z_\mathrm{ph}$.
It is evident that the significance of the dominant peak in the power spectrum
exceeds $5 \sigma$, however, given the uncertainty of such estimates
(see Section~\ref{sec:north}) we must limit ourselves
to the interval $\gtrsim (4 - 5) \sigma$ .

For comparison, Fig.~4b shows the power spectrum obtained by the
{Radon transform} method for the extended statistics
of clusters (26\,832) considering
photometric $z_\mathrm{ph}$. It is evident that
the peak at $k=k_\mathrm{max} \simeq 0.048 \pm 0.004~h$~Mpc$^{-1}$
(quasi-period $130 \pm 10~h^{-1}$~Mpc)
corresponds to the same significance
as the peak in Fig.~4a.
   
Thus, both methods, using only photometric redshifts $z_\mathrm{ph}$
of galaxy clusters in the southern hemisphere, give significant peaks
at close values of $k_\mathrm{max}$
in 1D power spectra,
calculated for the same direction $X_0$
close to the direction opposite to the
analogous direction $X_0$
in  the  northern hemisphere.  
  
\section{CONCLUSIONS AND DISCUSSIONS}
\label{sec:cd}

In this paper, based on the ``WHL12'' (Wen et al., 2012)
and ``WH22'' (Wen, Han, 2022) catalogs in the northern and southern hemispheres, 
respectively  (see also Section \ref{sec:data}), we consider
the possibility of the existence of an anisotropic quasi-periodic
structure (anomaly) in the spatial distribution of
cosmologically distant  ( $0.1 \leq z \leq 0.47$) galaxy clusters.
We use the previously proposed (Ryabinkov, Kaminker, 2021, 2024) 
method of projections
of the Cartesian coordinates of clusters onto different $X$ axes,
sequentially rotated within the boundaries of
certain regions in the sky (specified by observational data)
$-$ {Radon transform},
and its modification $-$ {scanning cuboid} method.
When the $X$ axis of the cuboid is rotated  its faces (edges, vertices)
are fixed in the $XYZ$ coordinate system rigidly connected to it,
and various samples of clusters (see Figs. 1a and 2a, b),
included in one of the two catalogs, fall into its volume.
This approach allows us to roughly localize
a possible anomaly in the accompanying  space.

In both variations of the one-dimensional projection method, 
after each rotation of the $X$ axis, the normalized distribution (\ref{NNX}) 
of the projections of the cluster coordinates onto 
this axis is calculated within the fixed interval
$X_1 \leq X \leq X_2$. Then the corresponding power spectrum (\ref{PXk}) 
is calculated  in the considered
interval $0.0 \leq k \leq 0.3$, while the cosmological redshifts
of the clusters  lie  in the interval $0.1 \leq z \leq 0.47$. 
Among all directions $X$ with different power spectra
we find (independently in each hemisphere)
the direction $X_0$ ($\alpha_0$ and $\delta_0$)
with the maximum peak height in the interval
$0.04 \lesssim k \lesssim 0.06$.

To  estimate  the significance of the  revealed  peaks, we use, 
following the works of Ryabinkov, Kaminker (2019, 2021), 
the exponential probability (\ref{calP}), (\ref{calF}) 
to obtain a given value of the peak amplitude in the power spectrum 
with a random distribution of the peak heights. 
The main element of this   estimation  is the calculation 
of the averaged power spectrum (\ref{PXk})
$\langle P(k) \rangle$ for all discrete $k$
from the interval under consideration.
Such averaging
was carried out based on the results of scanning two selected
regions (r1) and (r3), shown in Fig. 1b and 1c,
in the northern and southern hemispheres, respectively.

Let us summarize the  results  obtained. \\

1. In the northern and southern hemispheres, the presence 
of an anisotropic quasi-periodic
anomaly in the spatial distribution of
clusters with a characteristic scale in the range of
comoving  distances of $100 - 140~h^{-1}$~Mpc 
was independently detected. \\

2. The directions of $X_0\ (\alpha_0,\ \delta_0)$,
in which the amplitudes of the peaks in the power spectrum are maximum,
in both hemispheres are close to opposite ones:
$\alpha_0=170^\circ \pm 5^\circ, \ \delta_0=29^\circ \pm 5^\circ$
$-$ in the northern and $\alpha_0=346^\circ \pm 5^\circ, \
\delta_0 = -29^\circ \pm 5^\circ$ $-$ in the southern
hemispheres, respectively. \\

3. These features are revealed using both
spectroscopic $z_\mathrm{sp}$ (in the northern hemisphere)
and photometric redshifts $z_\mathrm{ph}$
(in the southern  hemisphere), determined
with an average error of $\delta z_\mathrm{ph} \lesssim 0.013$.
The significance of the maximum peaks in the power spectra
for the selected directions
at $k_\mathrm{max} \simeq (0.04 - 0.06)~h$~Mpc$^{-1}$
is\  $\gtrsim\  (4 - 5) \sigma$. \\

4. The strong dependence of the
significance of the quasi-periodic component on the accuracy of
photometric measurements of $\delta z_\mathrm{ph}$
opens up the possibility of additional control over the accuracy of
techniques  for determining of photometric $z_\mathrm{ph}$. \\

5. The total size of the anomaly along the  preferential   direction
roughly is $\sim 2500~h^{-1}$~Mpc.

It should be noted that the   estimate  of the significance 
of the peaks in the power spectrum
should be compared with the statistical  estimate  of the anisotropy
of the angular distribution of the peaks. 
Such an   estimate  gives much more modest 
significance levels $\sim (3 - 4)~\sigma$
of the   preferential  directions and,  probably, is a more realistic assessment
for the entire phenomenon of quasi-periodic anisotropy. 
A paper on this topic is being prepared
for publication.

Let us emphasize that to detect this anomaly, 
we use the method of projections onto a selected
$X$ axis, i.e. an integral method that accumulates information
from a large number of clusters distributed in space.
We assume that the anisotropic structure considered here
is very weakly manifested in the spatial
distribution of galaxy clusters,
in contrast to the much more noticeable quasi-regular formations
found in the distribution of superclusters at $z \lesssim 0.12$
(Einasto et al. 1997a,b, 2016;
Einasto 2014; Saar et al. 2002).
It is possible that the method of 1D projections used here
leads to a relative strengthening of the  sparse 
quasi-periodic component
and to an increase in its significance. 
It can be assumed that there is a connection between the features 
of the spatial distributions of galaxies and galaxy clusters, 
on the one side, and superclusters, on the other, 
although this assumption requires further research.

The possible existence of an anomaly, considered here, 
which extends  to  both hemispheres, 
may also be consistent with the results of a series of studies 
(Broadhurst et al. 1990; Szalay et al. 1993; Koo et al.  1993), 
in which a quasi-regular sequence of compressions  and 
rarefactions with a characteristic 
scale of  $\sim 130~h^{-1}$~Mpc  was discovered in 1D 
distributions of galaxies along a set of narrow cones in the direction of the
north and south galactic poles.
The total scale of such a quasi-regular inhomogeneity was $\sim 2000~h^{-1}$~Mpc.
Later simulations (Yoshida et al., 2001)
showed that the probability of random
origin of such 1D quasi-periodic
distributions is less than $10^{-3}$, which for equivalent
significance of the Gaussian processes  exceeds $3 \sigma$.

The pensil-beam distribution of galaxies differs
fundamentally   from the distribution
of projections of large arrays of objects (in this case $-$ galaxy clusters)
onto  the   preferential  $X_0$ axis. Nevertheless,
both types of  1D quasi-periodicity
can be    interrelated as two different ways
of probing one complex quasi-regular structure.
We only note that
a narrow beam of directions (Fig. 1b and 1c)
with a high level of significance of peaks 
in the power spectra
is  rotated by an angle of\  $\sim\  20^\circ - 30^\circ$ 
in right ascension $\alpha$ relative to the axis
connecting the north and south
galactic poles ($\alpha_\mathrm{ngp} = 192.86^\circ$, and
$\alpha_\mathrm{sgp} = 12.86^\circ\ (372.86^\circ)$).
This means that the directions of the $X_0$ axes
discussed in this paper differ significantly 
from the average direction of the narrow beams
oriented toward the galactic poles. 
On the other hand, the inclination   $\delta$ is close to the inclination of the
north ($\delta_\mathrm{ngp} = 27.13^\circ$) and south
($\delta_\mathrm{sgp} = - 27.13^\circ$) poles, respectively,
which is in better agreement with the hypothesis of the mutual connection of
these anomalies.

Note also that the
above-mentioned interval of quasi-periods includes the characteristic spatial
scales of $\sim (100 - 110)~h^{-1}$~Mpc,
corresponding to the main maximum of the baryon acoustic oscillations (BAO) 
in the power spectrum of matter density fluctuations in the Universe (see, e.g.,
Eisenstein et al.  2007, Alam et al.  2017).
It can be assumed that there is a connection between the BAO phenomenon and the
quasi-periods in the spatial distribution of galaxy clusters considered in this paper.
However, the attempt produced by Ryabinkov and  Kaminker (2019) 
to identify such a connection via modeling of the radial galaxy distributions
led to a low estimate of corresponding statistical significance. 
However, we continue to work on 3D modeling of a possible 
anisotropic distribution of the  BAO that would not contradict 
observations and  would be consistent with the results of this analysis.

In conclusion, it should be emphasized that the very existence 
of the considered anisotropic quasi-periodic anomaly, 
as well as its characteristics, remain hypothetical, 
requiring further confirmation and careful verification, 
including numerical  modeling.

\end{document}